# Where is liquid-vapor interface located in solutions?

Roumen Tsekov[1,2], Boiko Cohen[2], Boryan Radoev[2] and Hans J. Schulze[1]
[1]MPI Research Group for Colloids and Surfaces, TU Bergakademie, 09599 Freiberg, Germany
[2]Department Physical Chemistry, University of Sofia, 1164 Sofia, Bulgaria

Two-component liquid-vapor systems are modeled as two bulk phases divided by a two-dimensional surface phase and the mass and momentum balances are theoretically studied. Comparing the derived equations with some typical models of surface rheology, useful information about the interface location is obtained. It is demonstrated that the surface phase, set on the surface of tension, coincides with the equimolecular interface for insoluble surfactants, whereas it is placed on the surface of zero total mass density excess for soluble ones. The applicability of the model to surface electrostatics is also discussed by introduction of a two-dimensional Maxwell equation for the surface phase.

During the last decades much progress in the understanding of the transfer processes in heterogeneous systems was undoubtedly achieved but still there is no rigorous derivation of the phenomenological equations governing the mass, momentum and energy balances. The microscopic theories provide a description based on first principles. They are dealing, however, with complicated mathematical formalism and it is not possible until now to translate their exact results in the useful language of the macroscopic physics [1]. In contrast, the methods of non-equilibrium thermodynamics and hydrodynamics [2-5] are easier to handle and seem to be more appropriate for description of the problem. After Gibbs equilibrium heterogeneous systems are successfully described as constructed by homogeneous bulk and surface phases interchanging matter, momentum, energy, etc. In the literature [6, 7] the surface balances are presented in analogy with the bulk ones. An important point here is the location of the surface phase [8]. There is a number of dividing surfaces corresponding to different conservation laws, e.g. the surface of tension, equimolecular surface, etc., where the surface phase can be placed. The aim of the present paper is to show the location of the surface phase in some typical cases on the base of the existing experimental investigations of interfacial rheology.

The macroscopic system under consideration consists of a liquid being in equilibrium with its vapor. The mass density of the liquid, denoted by $\rho$, is much higher than the vapor mass density and for this reason, the latter is taken to be negligible. Let the liquid be a two-component solution with mass concentration $c$ of the second component and mass concentration $\rho - c$ of the solvent. Then the balances in the bulk should satisfy the well-known mass and momentum conservation laws [9]

$$\partial_t \rho + \nabla \cdot (\rho \mathbf{v}) = 0 \qquad \partial_t c + \nabla \cdot (c \mathbf{v} + \mathbf{J}) = 0 \qquad \partial_t (\rho \mathbf{v}) + \nabla \cdot (\rho \mathbf{v}\mathbf{v} + \mathbb{P}) = 0 \qquad (1)$$

where $\mathbf{v}$ is the hydrodynamic velocity, $\mathbf{J}$ is the diffusion flux, $\mathbb{P}$ is the pressure tensor, $\nabla$ is the nabla operator and $\partial_t \equiv \partial / \partial t$. The flux $\mathbf{J}$ is given by the Fick law

$$\mathbf{J} = -\rho D \nabla (c/\rho) \qquad (2)$$

expressed properly by the mass fraction [9]. In the case of a Newtonian liquid, the pressure tensor acquires the form

$$\mathbb{P} = p\mathbb{I} - \rho \upsilon \left[ (\nabla \mathbf{v}) + (\nabla \mathbf{v})^\mathrm{T} \right] - \rho (\kappa - 2\upsilon/3)(\nabla \cdot \mathbf{v})\mathbb{I} \qquad (3)$$

where $\mathbb{I}$ is the unit tensor and the superscript $^\mathrm{T}$ denotes a tensor-transpose. In general the diffusion coefficient $D$ and the kinematic shear $\upsilon$ and dilatational $\kappa$ viscosities are functions of the mass density and concentration. Assuming local equilibrium in the liquid, the pressure $p$ can also be expressed as a function of these quantities via the corresponding equation of state.

### Surface rheology

According to the Gibbs concept, the real system should be equivalent to two homogeneous bulk phases, liquid and vapor, divided by a two-dimensional homogeneous surface phase. Due to kinematic continuity of the system [8], the surface hydrodynamic velocity should coincide with the bulk velocity $\mathbf{v}$ taken at the surface phase plane. The laws of conservation at the surface have the same form as Eqs. (1) with non-zero right hand-sides accounting for the interaction with the bulk [2, 10]

$$\partial_t \gamma + \nabla_s \cdot (\gamma \mathbf{v}) = 0 \qquad \partial_t \Gamma + \nabla_s \cdot (\Gamma \mathbf{v} + \mathbf{J}_s) = \mathbf{J} \cdot \mathbf{n} \qquad \partial_t (\gamma \mathbf{v}) + \nabla_s \cdot (\gamma \mathbf{v}\mathbf{v} + \mathbb{P}_s) = \mathbb{P} \cdot \mathbf{n} \qquad (4)$$

Here $\gamma$ and $\Gamma$ are total and second component mass density excesses on the surface, $\mathbf{J}_s$ and $\mathbb{P}_s$ are interfacial diffusion flux and pressure tensor, $\mathbf{n}$ is the unit normal vector to the surface, $\nabla_s = \mathbb{I}_s \cdot \nabla$ and $\mathbb{I}_s = \mathbb{I} - \mathbf{nn}$ are the co-normal nabla operator and surface unite tensor, respectively. The right hand-sides of Eqs. (4) represent the bulk mass and momentum fluxes to the interface. In addition, the unit normal vector obeys the following equation [3, 10]

$$\partial_t \mathbf{n} + \nabla_s (\mathbf{n} \cdot \mathbf{v}) = 0 \tag{5}$$

Since the interface is a separate two-dimensional phase, the diffusion flux $\mathbf{J}_s$ and pressure tensor $\mathbb{P}_s$ should be similar to the bulk ones. Thus, the Fick law (2) for the surface phase reads

$$\mathbf{J}_s = -\gamma D_s \nabla_s (\Gamma / \gamma) \tag{6}$$

where $D_s$ is the surface diffusion coefficient. In the case of a Newtonian interfacial fluid, the surface pressure tensor $\mathbb{P}_s$ can be presented in analogy to Eq. (3) as

$$\mathbb{P}_s = -\sigma \mathbb{I}_s - \gamma \upsilon_s \left[ (\nabla_s \mathbf{v}) \cdot \mathbb{I}_s + \mathbb{I}_s \cdot (\nabla_s \mathbf{v})^T \right] - \gamma (\kappa_s - \upsilon_s)(\nabla_s \cdot \mathbf{v}) \mathbb{I}_s - \gamma \lambda_s (\nabla_s \mathbf{v}) \cdot \mathbf{nn} \tag{7}$$

where $\upsilon_s$ and $\kappa_s$ are the surface shear and dilatational kinematic viscosities, and the last term accounts for the transverse shear flow with corresponding kinematic viscosity $\lambda_s$. According to the usual prescriptions, the surface phase is placed at the surface of tension and for this reason the two-dimensional pressure in Eq. (7) is expressed by the surface tension $\sigma$. It is obvious that the surface diffusion coefficient and kinematic viscosities, having the same dimension with the bulk transport coefficients, depend on the mass density excesses and temperature.

The basic novelty in Eq. (7) is the presentation of the dynamic viscosities as products of the total mass density excess $\gamma$ and the corresponding kinematic viscosities. It is similar to the bulk but the physical meaning here is more essential. This presentation accounts for the necessary condition the total mass density excess and the dynamic viscosities to have one and the same sign which is required by the positive entropy production in the system (remember that the mass density excess could be either positive or negative). Hence, the surface kinematic vis-cosities are to be positive. Moreover, they have the same dimension with the bulk ones which reflects in an easier way for juxtaposition and modeling.

## Linear hydrodynamics

The general description of the coupled bulk and interfacial transports via Eqs. (1-7) is complicated. A well-known simplification is the model of the incompressible liquid. Further, in the case of small deviations from equilibrium, the non-linear terms in the conservation balances drop out, Eqs. (1-3) reduce to

$$\nabla \cdot \mathbf{v} = 0 \qquad \partial_t c = D\Delta c \qquad \rho \partial_t \mathbf{v} = -\nabla p + \rho \upsilon \Delta \mathbf{v} \qquad (8)$$

and Eqs. (4-7) acquire the forms

$$\partial_t \gamma + \gamma \nabla_s \cdot \mathbf{v} = 0$$
$$\partial_t \Gamma + \Gamma \nabla_s \cdot \mathbf{v} = D_s \Delta_s \Gamma - (\Gamma/\gamma) D_s \Delta_s \gamma - D \nabla_n c$$
$$\gamma \partial_t \mathbf{v}_s = \nabla_s \sigma + \gamma \upsilon_s \Delta_s \mathbf{v}_s + \gamma \kappa_s \nabla_s \nabla_s \cdot \mathbf{v} - \rho \upsilon (\nabla_s v_n + \nabla_n \mathbf{v}_s) \qquad (9)$$
$$\gamma \partial_t v_n = -\sigma \nabla_s \cdot \mathbf{n} + \gamma \lambda_s \Delta_s v_n + p - 2\rho \upsilon \nabla_n v_n$$
$$\partial_t \mathbf{n} + \nabla_s v_n = 0$$

where $\Delta_s \equiv \nabla_s^2$ and $\mathbf{v}_s = \mathbb{I}_s \cdot \mathbf{v}$. Equations (9) represent the boundary conditions to Eqs. (8).

Let us consider first the case of a pure solvent, i.e. $c = 0$ and $\Gamma = 0$. In this case the boundary conditions to Eqs. (8) are usually described by the following equations [2]

$$\nabla_s v_n + \nabla_n \mathbf{v}_s = 0 \qquad \sigma \Delta_s \zeta + p - 2\rho \upsilon \nabla_n v_n = 0 \qquad \partial_t \zeta = v_n$$

where $\zeta$ is the normal deformation of the surface. These equations will coincide with Eqs. (9) if $\gamma = 0$. Therefore, the surface phase in one-component systems is placed at the surface of tension determined by the elastic force and momentum balances and at the equimolecular surface where the surface mass density excess of the solvent is zero. Rigorous calculations show that these two surfaces do not coincide in general. However, the differences are small (a few Å) and for this reason, they are not important for macroscopic hydrodynamics [4].

For two-component systems, there are two distinguished models of surface rheology reported in the literature. The first one is usually applied to insoluble monolayers $(c = 0)$. This case is well described by the following boundary conditions

$$\partial_t \Gamma + \Gamma \nabla_s \cdot \mathbf{v} = 0$$
$$\Gamma \partial_t \mathbf{v}_s = (\partial_\Gamma \sigma) \nabla_s \Gamma + \Gamma \upsilon_s \Delta_s \mathbf{v}_s + \Gamma \kappa_s \nabla_s \nabla_s \cdot \mathbf{v} - \rho \upsilon (\nabla_s v_n + \nabla_n \mathbf{v}_s) \quad (10)$$
$$\Gamma \partial_t v_n = \sigma \Delta_s \zeta + \Gamma \lambda_s \Delta_s v_n + p - 2\rho \upsilon \nabla_n v_n$$
$$\partial_t \zeta = v_n$$

The main feature of these balances is the absence of the surface diffusion which is experimentally confirmed [11]. As is seen, Eqs. (10) are a particular case of Eqs. (9) with $\gamma = \Gamma$. This means that the surface phase coincides again with the solvent equimolecular dividing surface. The result that the mass density excess of the solvent is zero shows that insoluble surfactants do not change the place of the surface phase of the pure solvent. They only concentrate there and thus indicate the place of the solvent surface phase. Note that the lack of the surface diffusion is due to the fact that the surface phase is one-component, and not because $D_s = 0$.

The usual boundary balances for the well-soluble surfactants are [2]

$$\partial_t \Gamma + \Gamma \nabla_s \cdot \mathbf{v} = D_s \Delta_s \Gamma - D \nabla_n c$$
$$(\partial_\Gamma \sigma) \nabla_s \Gamma = \rho \upsilon (\nabla_s v_n + \nabla_n \mathbf{v}_s) \quad (11)$$
$$\sigma \Delta_s \zeta + p - 2\rho \upsilon \nabla_n v_n = 0$$
$$\partial_t \zeta = v_n$$

In contrast to the previous case, they depend substantially on the surface diffusion but the effect of the surface viscous flow is absent. This is again in accordance with experimental results [12]. As is seen, Eqs. (11) follow from the general balances (9) when $\gamma = 0$. Therefore, the surface phase of soluble surfactants coincides with the tension and zero total mass density excess surfaces. This means that the solutions have their own location of the surface phase different from that of the pure solvent. As in the bulk, the surface dynamic viscosities are presented in Eq. (7) as products of the corresponding kinematic viscosities and $\gamma$. For this reason they disappear in the case of the well-soluble surfactants since $\gamma = 0$.

## Electrostatics

An interesting discrimination between the two models discussed above appears if one takes into account the influence of electrostatics. Supposing the considered liquid is a dielectric substance, the static electric field intensity $\mathbf{E}$ should satisfy the well-known Maxwell equations [13]

$$\nabla \cdot (\varepsilon \mathbf{E}) = 0 \tag{12}$$

$$\nabla \times \mathbf{E} = 0 \tag{13}$$

where $\varepsilon$ is the dielectric permittivity of the liquid. Since the contribution of the electric field to diffusion and hydrodynamics are quadratic [14], in the frames of the linear theory, the equations obtained in the previous section will not be affected by the electrostatics.

To solve completely the electrostatic problem one needs boundary conditions on the liquid-vapor interface. It is well-known that Eq. (13) leads to equivalence of the tangential components of the field on the bulk phase dividing surface. Hence, following our concept we can introduce a two-dimensional Maxwell equation on the surface corresponding to Eq. (12)

$$\nabla_s \cdot (\varepsilon_s \mathbf{E}_s) = \varepsilon E_n - E_n^0 \tag{14}$$

where $\varepsilon_s$ is the surface dielectric permittivity and $E_n^0$, being the normal field intensity in vacuum, stands here for the vapor phase. Equation (14) relates the surface projections of the electric field and in this way completes Eq. (12) by the desired boundary condition.

Following the natural dependence of the dielectric permittivity on the liquid density, $\varepsilon = 1 + \rho\alpha$ [13], one can propose a similar relation $\varepsilon_s = \gamma\alpha_s$ at the surface. Since $\varepsilon_s$ turns to be an excess quantity, it vanishes at $\gamma = 0$ and there is no vacuum surface dielectric property. Therefore, for pure liquids and their mixtures with soluble surfactants, Eq. (14) reduces to the classical result $\varepsilon E_n = E_n^0$. More interesting is the case of insoluble surfactants where $\gamma = \Gamma$. In this case, the linearized electrostatic boundary condition reads

$$\varepsilon E_n + \Gamma \alpha_s \nabla_n E_n = E_n^0 \tag{15}$$

This equation exhibits a more complicated relation between the normal projections of the field due to specific properties of the surface layer of surfactant molecules.

An additional effect could arise from the existence of a surface dipole moment per unit area $\mathbf{d}$ of the interfacial layers [14] which certainly behave like liquid crystals. Note that the electric field generated by $\mathbf{d}$ is purely fluctuational and its mean value is zero. In this case, the surface Maxwell equation takes the form $\varepsilon E_n - \nabla_s \cdot (\varepsilon_s \mathbf{E}_s + 4\pi \mathbf{d}) = E_n^0$. For insoluble surfactants the co-normal divergence of the surface dipole moment could be presented in the linear case by $\nabla_s \cdot \mathbf{d} = (\partial_\Gamma \mathbf{d}) \cdot \nabla_s \Gamma - d_n \Delta_s \zeta$ and thus it is connected to the interfacial diffusion and hydrodynamics. In a previous paper [15] the effect of the surface dipole moment and permittivity on the interfacial waves has been theoretically described and a criterion for stability is obtained.

## Conclusions

In the present paper, the mass and momentum balances in a two-component liquid-vapor system are studied. Comparing the equations derived with some typical rheological models, useful information about the location of the interface is obtained. It was demonstrated that the surface phase for insoluble surfactants coincides with the equimolecular interface of the solvent while for soluble ones it is placed on the surface of zero total mass density excess. An usual delusion is that there is no big difference between the locations of these two dividing surfaces defined by $\gamma = \Gamma$ and $\gamma = 0$. From an elementary mass balance, however, the distance between them is equal to $\Gamma / \rho$ and there is not restriction for this quantity to be of macroscopic order. The main question solved in the present paper is which one of these two surfaces is closer to the real surface or in other words what is the value of $\gamma$ on the surface of tension. In a previous work [8], surface wave dispersion relations have been derived for both the models reported here and the applicability of light scattering measurements for experimental discrimination of the surface diffusion and viscosity effects has been demonstrated.

The two separate cases discussed here demonstrate two particular kinds of relaxation behavior of the surface processes. In general, surface rheology is described neither by Eqs. (10) nor by Eqs. (11). It satisfies Eqs. (9). Moreover, the Gibbs approach to interfacial rheology is applicable to fluids which molecules are smaller than the transition depth of the interfacial region [4]. For this reason, the present theory may not work for macromolecular surfactants. In the latter case, additional complications arise from usually the non-Newtonian surface rheology of the second component.